# Reply to the Comment on "Negative Landau damping in bilayer graphene"

Tiago A. Morgado[1], Mário G. Silveirinha[1,2]

[1]*Instituto de Telecomunicações and Department of Electrical Engineering, University of Coimbra, 3030-290 Coimbra, Portugal*

[2]*University of Lisbon, Instituto Superior Técnico, Avenida Rovisco Pais, 1, 1049-001 Lisboa, Portugal*

*E-mail:* tiago.morgado@co.it.pt, mario.silveirinha@co.it.pt

In the preceding Comment [1] Svintsov and Ryzhii (SR) criticize the conductivity derived in our Letter using the self-consistent field (SCF) approach [2, 3]

$$\sigma_{\text{g}}^{\text{drift}}(\omega, q_x) = (\omega/\tilde{\omega}) \sigma_{\text{g}}(\tilde{\omega}, q_x). \quad (1)$$

Here, $\sigma_{\text{g}}(\omega, q_x)$ is the nonlocal conductivity with no drift and $\tilde{\omega} = \omega - q_x v_0$. The drift effect was modeled by the interaction Hamiltonian $\hat{H}_{\text{int,drift}} = \mathbf{v}_0 \cdot \hat{\mathbf{p}}$, $\mathbf{v}_0 = v_0 \hat{\mathbf{x}}$ being the drift velocity and $\hat{\mathbf{p}} = -i\hbar \nabla$ [2]. The use of this interaction Hamiltonian was motivated by an analogy with moving media [4]. Equation (1) extends to graphene the well-known result $\varepsilon^{\text{drift}}(\omega, q_x) = \varepsilon(\omega - q_x v_0, q_x)$ for a drift-biased plasma [5-7] (for 3D materials $\varepsilon(\omega) = 1 + \sigma/(-i\omega\varepsilon_0)$).

SR argue that because the electrons in graphene are massless the Galilean Doppler shift cannot be used. They rely on the distribution $f_{\text{drift}}(\mathbf{k}) = f^0(\mathcal{E}_{\mathbf{k}}^0 - \hbar \mathbf{k} \cdot \mathbf{v}_0)$, which is applicable when the electron-electron (e-e) scattering predominates [8]. Here, $\mathcal{E}_{\mathbf{k}}^0$ is the energy dispersion of the relevant electronic band and $f^0(\mathcal{E})$ is the Fermi-Dirac distribution. SR use $f_{\text{drift}}(\mathbf{k})$ in the Lindhard formula. However, they miss a subtle point. In the shifted Fermi distribution $\hbar\mathbf{k}$ is a *kinetic* momentum rather than a *canonical* momentum (see Ref. [9, App. H]). The canonical momentum is $\mathbf{p} = \hbar\mathbf{k} - e\mathbf{A}$ with $\mathbf{A}$ the vector potential due to the static electric field $\mathbf{E}_0 = E_0 \hat{\mathbf{x}}$. The vector potential is $\mathbf{A}(t) = -(t-t_0)\mathbf{E}_0$ in the intervals between e-e collisions ($t=t_0$ is the time instant of a collision).

The Lindhard formalism relies on the time evolution of Bloch states ($\psi_{n\kappa}$). The Bloch wave vector $\kappa$ determines the canonical momentum. This means that the relevant distribution for the Lindhard formula is a canonical momentum distribution [10]. It is roughly $\tilde{f}_{\text{drift}}(\kappa) \approx f_{\text{drift}}(\kappa + \hbar^{-1}e\mathbf{A}) \approx f^0(\mathcal{E}_\kappa^0 + e\langle\mathbf{A}\rangle\cdot\mathbf{v}_\kappa^0 - \hbar\kappa\cdot\mathbf{v}_0)$ where $\mathbf{v}_\kappa^0 = \hbar^{-1}\partial_\kappa \mathcal{E}_\kappa^0$. In the second identity, we used a Taylor expansion, replaced $\mathbf{A}(t)$ by its time average $\langle\mathbf{A}\rangle$, and dropped the term $e\mathbf{A}\cdot\mathbf{v}_0$ because it is of second order ($\sim E_0^2$). Moreover, since $\mathbf{E}_0 = E_0 \hat{\mathbf{x}}$ is space independent the canonical momentum of an electron must be preserved by the static field (it is also preserved by the e-e collisions as on average they are independent of the space coordinates). This implies that $\tilde{f}_{\text{drift}}(\kappa) = f^0(\mathcal{E}_\kappa^0)$. Thus, when the e-e collisions predominate one must have $e\langle\mathbf{A}\rangle\cdot\mathbf{v}_\kappa^0 = \hbar\kappa\cdot\mathbf{v}_0$.

Substitution of $\tilde{f}_{\text{drift}}(\kappa) = f^0(\mathcal{E}_\kappa^0)$ in the Lindhard formula yields (taking the band overlap integral $F_{\kappa,\kappa+\mathbf{q}} \approx 1$):

$$\sigma_{\omega,q}^{\text{drift}} = \frac{i\omega e^2}{q^2}\frac{g_s g_v}{(2\pi)^2}\iint d^2\kappa \frac{f^0(\mathcal{E}_\kappa^0) - f^0(\mathcal{E}_{\kappa+\mathbf{q}}^0)}{\hbar\omega + \mathcal{E}_\kappa - \mathcal{E}_{\kappa+\mathbf{q}}}. \tag{2}$$

Here, $\mathcal{E}_\kappa \approx \langle\psi_\kappa|\hat{H}_0(\hat{\mathbf{p}} + e\langle\mathbf{A}\rangle)|\psi_\kappa\rangle \approx \mathcal{E}_\kappa^0 + \langle e\mathbf{A}\rangle\cdot\langle\psi_\kappa|\partial_\mathbf{p}\hat{H}_0(\hat{\mathbf{p}})|\psi_\kappa\rangle$ is the average electron energy during the interaction with the static field. Combining $\langle\psi_\kappa|\partial_\mathbf{p}\hat{H}_0(\hat{\mathbf{p}})|\psi_\kappa\rangle = \mathbf{v}_\kappa^0$ and $e\langle\mathbf{A}\rangle\cdot\mathbf{v}_\kappa^0 = \hbar\kappa\cdot\mathbf{v}_0$, it is found that $\mathcal{E}_\kappa \approx \mathcal{E}_\kappa^0 + \hbar\kappa\cdot\mathbf{v}_0$. Note that $\mathcal{E}_\kappa \neq \mathcal{E}_\kappa^0$ because the electron is accelerated by $\mathbf{E}_0$. Substituting $\mathcal{E}_\kappa \approx \mathcal{E}_\kappa^0 + \hbar\kappa\cdot\mathbf{v}_0$ in Eq. (2), we recover Eq. (1) and the Galilean Doppler-shift (see [10] for additional discussion and a derivation with the Boltzmann equation).

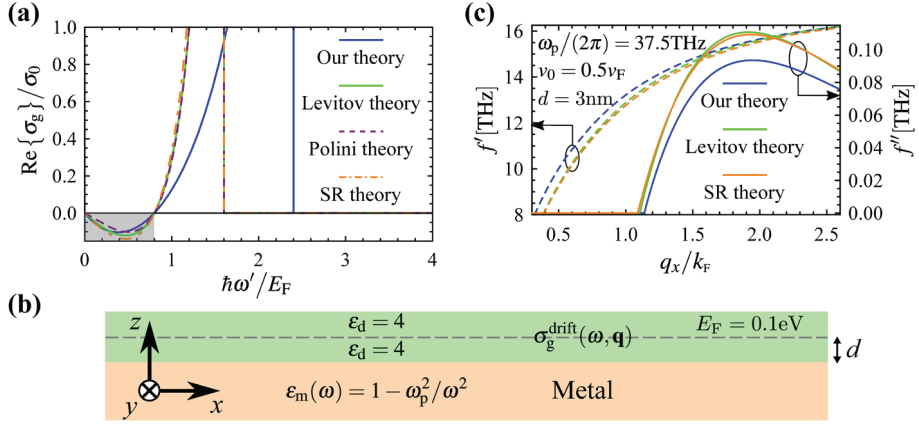

**Fig. 1.** (a) $\text{Re}\{\sigma_g^{\text{drift}}\}$ in the UHP as a function of $\hbar\omega'/E_F$ for $q_x = 1.6k_F$, $E_F = 0.1$ eV, $v_0 = 0.5v_F$, and $\omega'' = 0^+$. The NLD region is shaded in gray. (b) A drift-current biased graphene sheet and a metal half-space are separated by the distance $d$ with a dielectric; (c) $\omega/(2\pi) = f' + if''$ for the unstable mode as a function of $q_x$.

Regarding the second point raised by SR about the long-wavelength approximation, we underline that the nonlocality neither precludes the negative Landau damping (NLD) nor the emergence of instabilities in graphene platforms. Indeed, the square root singularity of $\sigma_g$ is compatible with gain regimes because when $\tilde{\omega} = \omega - q_x v_0$ is negative the pre-factor $\omega/\tilde{\omega}$ of (1) is also negative. Thus, the NLD ($\text{Re}\{\sigma_g^{\text{drift}}(\omega, q_x)\} < 0$) can occur in the real-frequency axis (Fig. 1a), or in the upper-half frequency plane (UHP), $\omega'' = \text{Im}\{\omega\} \geq 0$ with $\omega = \omega' + i\omega''$ [10, 11]. The same result is predicted by the collisionless SR, Levitov's [12] and Polini's [13] models (see Fig. 1a). Thus, similar to Ref. [2], by coupling the drift-current biased graphene to a resonant system (here a metal half-space) (Fig. 1b) it is possible to spontaneously generate THZ and IR radiation (Fig. 1c). The collisionless models of SR and Levitov predict quantitatively similar unstable regimes; all the models predict solutions with $f'' = \omega''/(2\pi) > 0$ that grow exponentially with time.

In summary, the drift-biased conductivity is ruled by a Galilean transformation when the e-e collisions force the electron gas to move with a constant velocity. The instabilities predicted in our Letter may be observed in properly designed drift-current biased graphene platforms.

Acknowledgments: This work was partially funded by Fundação para Ciência e a Tecnologia under Project PTDC/EEITEL/4543/2014 and by Instituto de Telecomunicações under Project No. UID/EEA/50008/2019. T. A. Morgado acknowledges financial support by FCT under the CEEC Individual 2017 contract as assistant researcher with reference CT/Nº004/2019-F00069 established with IT – Coimbra.

# Supplementary Information

## *A. Further discussion on the distribution function and Lindhard formula*

The Lindhard formalism relies on the time evolution (described with perturbation theory) of the crystal Bloch states ($\psi_{n\kappa}$) under the influence of an electric field. The time evolution of the wave function is ruled by:

$$i\hbar \partial_t \psi = \left[ \hat{H}_0 \left( -i\hbar \nabla + e\mathbf{A}, \mathbf{r} \right) - e\phi \right] \psi . \tag{S1}$$

Here, $\phi, \mathbf{A}$ are the scalar and the vector potentials. As explained in the main text, the static electric field is modeled by a *space independent* vector potential. The dynamic (and longitudinal) electric field (with space and time variation of the form $e^{iq_x x} e^{-i\omega t}$) is described by $\phi$.

Let us first ignore the dynamic field ($\phi \approx 0$) which is assumed weak compared to the static bias. Then, for a initial state of the form $\psi(\mathbf{r}, t = 0) = \psi_{n\kappa}$ (where $\psi_{n\kappa}$ is a Bloch eigenstate of $\hat{H}_0(-i\hbar\nabla, \mathbf{r})$) the time evolution determined by Eq. (S1) preserves the Bloch wave vector (this is so because $\mathbf{A}$ depends exclusively on time). In other words, for any time $t$ the function $\psi(\mathbf{r}, t) e^{-i\kappa \cdot \mathbf{r}}$ is periodic in space. Thus, the crystal momentum ($\kappa$) is preserved by the interaction with the static bias. Due to this reason the canonical momentum distribution is simply $\tilde{f}_{\text{drift}}(\kappa) = f^0\left(\mathcal{E}_\kappa^0\right)$, as it is unaffected by the static bias. In contrast, the kinetic momentum distribution is affected by the static bias as $\hbar \mathbf{k} = \hbar \kappa + e\langle \mathbf{A} \rangle_\kappa$. This leads to a shifted (in the direction of the drift) (kinetic) Fermi distribution $f_{\text{drift}}$. Notice that the time averaged vector potential $\langle \mathbf{A} \rangle_\kappa$ in general may depend on the canonical momentum.

Since the Lindhard formula gives the conductivity in terms of a sum (integral) of the contributions of the electronic states, and since each state is parameterized by the crystal (canonical) momentum, it is evident that the integral must be over $\hbar\boldsymbol{\kappa}$ rather than over $\hbar\mathbf{k}$ (kinetic momentum). Thus, the Lindhard formula must be written as in Eq. (2) of the main text, repeated here for convenience:

$$\sigma_{\omega,q}^{\text{drift}} = \frac{i\omega e^2}{q^2} \frac{g_s g_v}{(2\pi)^2} \iint d^2\boldsymbol{\kappa} \frac{\tilde{f}_{\text{drift}}(\boldsymbol{\kappa}) - \tilde{f}_{\text{drift}}(\boldsymbol{\kappa}+\mathbf{q})}{\hbar\omega + \mathcal{E}_{\boldsymbol{\kappa}} - \mathcal{E}_{\boldsymbol{\kappa}+\mathbf{q}}}, \quad (S2)$$

where $\mathcal{E}_{\boldsymbol{\kappa}} \approx \mathcal{E}_{\boldsymbol{\kappa}}^0 + \hbar\boldsymbol{\kappa}\cdot\mathbf{v}_0$ and $\tilde{f}_{\text{drift}}(\boldsymbol{\kappa}) = f^0(\mathcal{E}_{\boldsymbol{\kappa}}^0)$. In contrast the formula used in the derivation of SR is:

$$\sigma_{\omega,q}^{\text{SR}} = \frac{i\omega e^2}{q^2} \frac{g_s g_v}{(2\pi)^2} \iint d^2\mathbf{k} \frac{f_{\text{drift}}(\mathbf{k}) - f_{\text{drift}}(\mathbf{k}+\mathbf{q})}{\hbar\omega + \mathcal{E}_{\mathbf{k}}^0 - \mathcal{E}_{\mathbf{k}+\mathbf{q}}^0}. \quad (S3)$$

It is relevant to note that since $e\langle\mathbf{A}\rangle\cdot\mathbf{v}_{\boldsymbol{\kappa}}^0 = \hbar\boldsymbol{\kappa}\cdot\mathbf{v}_0$ and $\hbar\mathbf{k} = \hbar\boldsymbol{\kappa} + e\langle\mathbf{A}\rangle$, the mean electron energy ($\mathcal{E}_{\boldsymbol{\kappa}}$) can be expressed in terms of the kinetic momentum as $\mathcal{E}_{\boldsymbol{\kappa}} \approx \mathcal{E}_{\mathbf{k}}^0$ (notice the different indices). Moreover, it is possible to write $f_{\text{drift}}(\mathbf{k}) = f^0(\mathcal{E}_{\mathbf{k}}^0 - \hbar\mathbf{k}\cdot\mathbf{v}_0) \approx f^0(\mathcal{E}_{\boldsymbol{\kappa}}^0 + e\langle\mathbf{A}\rangle\cdot\mathbf{v}_{\boldsymbol{\kappa}}^0 - \hbar\boldsymbol{\kappa}\cdot\mathbf{v}_0) \approx f^0(\mathcal{E}_{\boldsymbol{\kappa}}^0) = \tilde{f}_{\text{drift}}(\boldsymbol{\kappa})$, where we used again $e\langle\mathbf{A}\rangle\cdot\mathbf{v}_{\boldsymbol{\kappa}}^0 = \hbar\boldsymbol{\kappa}\cdot\mathbf{v}_0$ and $\hbar\mathbf{k} = \hbar\boldsymbol{\kappa} + e\langle\mathbf{A}\rangle$, and ignored terms that are of second order in the field interaction.

The main reason why Eqs. (S2)-(S3) predict very different results is that the identities $\mathcal{E}_{\boldsymbol{\kappa}} \approx \mathcal{E}_{\mathbf{k}}^0$ and $f_{\text{drift}}(\mathbf{k}) \approx \tilde{f}_{\text{drift}}(\boldsymbol{\kappa})$ discussed in the previous paragraph do *not* imply that $\mathcal{E}_{\boldsymbol{\kappa}+\mathbf{q}} \approx \mathcal{E}_{\mathbf{k}+\mathbf{q}}^0$ and $f_{\text{drift}}(\mathbf{k}+\mathbf{q}) \approx \tilde{f}_{\text{drift}}(\boldsymbol{\kappa}+\mathbf{q})$ because the relation between the kinetic and canonical momentum is *nonlinear* ($\hbar\mathbf{k} = \hbar\boldsymbol{\kappa} + e\langle\mathbf{A}\rangle_{\boldsymbol{\kappa}}$; again we underline that $\langle\mathbf{A}\rangle_{\boldsymbol{\kappa}}$ generally depends on the canonical momentum). This is where Eq. (S3) goes wrong. Note that the effect of the dynamical field $\phi \sim e^{iq_x x} e^{-i\omega t}$ in Eq. (S1) is to create a weak perturbation of the wave function that has canonical momentum $\boldsymbol{\kappa}+\mathbf{q}$. Thus $\sigma_{\omega,q}^{\text{drift}}$ must unquestionably depend on the interactions between states with canonical momentum $\hbar\boldsymbol{\kappa}$ and $\hbar(\boldsymbol{\kappa}+\mathbf{q})$, consistent with the term $\tilde{f}_{\text{drift}}(\boldsymbol{\kappa}) - \tilde{f}_{\text{drift}}(\boldsymbol{\kappa}+\mathbf{q})$ in the integrand of Eq. (S1). In contrast, the formula used by SR predicts the interactions between states with *kinetic* momentum $\hbar\mathbf{k}$ and $\hbar(\mathbf{k}+\mathbf{q})$ (because of the term $f_{\text{drift}}(\mathbf{k}) - f_{\text{drift}}(\mathbf{k}+\mathbf{q})$), which has no physical basis.

## *B. Derivation of the graphene conductivity with the Boltzmann theory*

The Boltzmann equation predicts that the electron distribution function $f = f(\mathbf{r},\mathbf{p},t)$ satisfies (we use a distribution function $f$ such that $\mathbf{p}$ represents the canonical momentum rather than the kinetic momentum, $\mathbf{p}+e\mathbf{A}$):

$$\partial_t f + \dot{\mathbf{r}}\cdot(\partial f/\partial \mathbf{r}) + \dot{\mathbf{p}}\cdot(\partial f/\partial \mathbf{p}) = (\partial f/\partial t)_{\text{coll}}. \tag{S4}$$

The semiclassical Hamiltonian is determined by $\mathcal{H}_{\text{sc}}(\mathbf{r},\mathbf{p}) = \mathcal{E}^0_{\hbar^{-1}(\mathbf{p}+e\mathbf{A})} - e\phi$ with $\phi, \mathbf{A}$ the scalar and the vector potentials [R1, Appendix H] ($\mathbf{E} = -\nabla\phi - \partial_t\mathbf{A}$). By definition $\dot{\mathbf{r}} = \partial\mathcal{H}_{\text{sc}}/\partial\mathbf{p}$ and $\dot{\mathbf{p}} = -\partial\mathcal{H}_{\text{sc}}/\partial\mathbf{r}$. Here, $\mathcal{E}^0_{\mathbf{k}}$ determines the energy dispersion of the relevant electronic band.

The collisions with the ionic lattice are modeled with the relaxation time approximation $(\partial f/\partial t)_{\text{coll}} = (f^0(\mathcal{E}^0_{\boldsymbol{\kappa}}) - f)/\tau_{\text{ion}}$ where $f^0(\mathcal{E})$ is the Fermi-Dirac distribution. On the other hand, the electron-electron (e-e) collisions are modeled by a periodic in time (e.g., with a saw-tooth type profile) vector potential, such that $\mathbf{A}(t) = -(t-t_i)\mathbf{E}_0$ in the generic interval $t_i^+ < t < t_{i+1}^-$ with $t = t_i$ the time instants of the e-e collisions which may depend on $\mathbf{p}$ (thereby $\mathbf{A} = \mathbf{A}(\mathbf{p},t)$). Here, $\mathbf{E}_0 = E_0\hat{\mathbf{x}}$ is the static electric field. The time-averaged $\mathbf{A}$ is $\langle\mathbf{A}\rangle = -\tau_{\boldsymbol{\kappa}}\mathbf{E}_0$ where $\tau_{\boldsymbol{\kappa}}$ determines the e-e scattering time and $\boldsymbol{\kappa} = \hbar^{-1}\mathbf{p}$. It is assumed that $\tau_{\boldsymbol{\kappa}} \ll \tau_{\text{ion}}$ so that the e-e collisions predominate. The scalar potential determines the dynamic (longitudinal) electric field $-\nabla\phi = \mathbf{E}_1 e^{iq_x x} e^{-i\omega t}$.

The solution of the Boltzmann equation without the dynamic field ($\phi = 0$) is exactly $f = f^0\left(\mathcal{E}_{\boldsymbol{\kappa}}^0\right)$. With the dynamic field, the solution can be written as $f = f^0 + f^1$ with $f^1$ satisfying:

$$\partial_t f^1 + \partial_{\mathbf{p}}\left[\mathcal{E}_{\hbar^{-1}(\mathbf{p}+e\mathbf{A})}^0\right] \cdot \left(\partial f^1 / \partial \mathbf{r}\right) + \left(-e\mathbf{E}_1 e^{iq_x x} e^{-i\omega t}\right) \cdot \left(\partial f^0 / \partial \mathbf{p} + \partial f^1 / \partial \mathbf{p}\right) = -f^1 / \tau_{\text{ion}}. \quad (S5)$$

In a linear response approximation, the term $\partial f^1 / \partial \mathbf{p}$ must be dropped. Using $\mathcal{E}_{\hbar^{-1}(\mathbf{p}+e\mathbf{A})}^0 \approx \mathcal{E}_{\hbar^{-1}\mathbf{p}}^0 + e\mathbf{A} \cdot \mathbf{v}_{\hbar^{-1}\mathbf{p}}^0$ one can write $\dot{\mathbf{r}} = \partial_{\mathbf{p}}\left[\mathcal{E}_{\hbar^{-1}(\mathbf{p}+e\mathbf{A})}^0\right] \approx \partial_{\mathbf{p}}\left[\mathcal{E}_{\hbar^{-1}\mathbf{p}}^0 + e\mathbf{A} \cdot \mathbf{v}_{\hbar^{-1}\mathbf{p}}^0\right]$ where $\mathbf{v}_{\boldsymbol{\kappa}}^0 = \hbar^{-1}\partial_{\boldsymbol{\kappa}} \mathcal{E}_{\boldsymbol{\kappa}}^0$. The function $\partial_{\mathbf{p}}\left[\mathcal{E}_{\hbar^{-1}\mathbf{p}}^0 + e\mathbf{A} \cdot \mathbf{v}_{\hbar^{-1}\mathbf{p}}^0\right]$ varies periodically in time (because the vector potential also does) and its time-average is $\langle \dot{\mathbf{r}} \rangle \approx \partial_{\mathbf{p}}\left[\mathcal{E}_{\boldsymbol{\kappa}}^0 - e\mathbf{E}_0 \cdot \mathbf{v}_{\boldsymbol{\kappa}}^0 \tau_{\boldsymbol{\kappa}}\right] = \mathbf{v}_{\boldsymbol{\kappa}}^0 + \delta\mathbf{v}_{\boldsymbol{\kappa}}$ with $\delta\mathbf{v}_{\boldsymbol{\kappa}} = -e\hbar^{-1}\partial_{\boldsymbol{\kappa}}\left(\mathbf{E}_0 \cdot \mathbf{v}_{\boldsymbol{\kappa}}^0 \tau_{\boldsymbol{\kappa}}\right)$. Replacing $\dot{\mathbf{r}}$ by $\langle \dot{\mathbf{r}} \rangle$ in Eq. (S5) we get:

$$\left(\tau_{\text{ion}}^{-1} + \partial_t\right)f^1 + \left(\mathbf{v}_{\boldsymbol{\kappa}}^0 + \delta\mathbf{v}_{\boldsymbol{\kappa}}\right) \cdot \left(\partial f^1 / \partial \mathbf{r}\right) = e\mathbf{E}_1 e^{iq_x x} e^{-i\omega t} \cdot \partial f^0 / \partial \mathbf{p}. \quad (S6)$$

The solution of this equation is

$$f^1 = \frac{eE_1 e^{iq_x x} e^{-i\omega t}}{\tau_{\text{ion}}^{-1} - i\left(\omega - v_{x,\boldsymbol{\kappa}}^0 q_x - \delta v_{x,\boldsymbol{\kappa}} q_x\right)} \frac{\partial f^0}{\partial p_x}$$
$$= \frac{eE_1 e^{iq_x x} e^{-i\omega t}}{\tau_{\text{ion}}^{-1} - i\left(\omega - v_{x,\boldsymbol{\kappa}}^0 q_x - \delta v_{x,\boldsymbol{\kappa}} q_x\right)} \frac{\partial f^0\left(\mathcal{E}_{\boldsymbol{\kappa}}^0\right)}{\partial \mathcal{E}} v_{x,\boldsymbol{\kappa}}^0. \quad (S7)$$

The impact of the static electric field on the dynamic response depends critically on $\tau_{\boldsymbol{\kappa}}$, i.e., on how the scattering time varies with the canonical momentum. If the e-e collisions act to create a "drift", i.e., a coherent motion of the whole electron gas with a constant velocity, then the velocity $\delta\mathbf{v}_{\boldsymbol{\kappa}}$ must be independent of $\boldsymbol{\kappa}$. In the graphene case, this situation can occur when $\tau_{\boldsymbol{\kappa}} \sim |\boldsymbol{\kappa}| \sim \mathcal{E}_{\boldsymbol{\kappa}}^0$, i.e., when the collision time is proportional to the energy for states near the Fermi level. In such a situation, $\mathbf{v}_{\boldsymbol{\kappa}}^0 \tau_{\boldsymbol{\kappa}} = v_F \hat{\boldsymbol{\kappa}} \tau_{\boldsymbol{\kappa}} \sim \boldsymbol{\kappa}$ and therefore $\delta v_{x,\mathbf{k}}$ is of the form $\delta v_{x,\mathbf{k}} = v_0$, with $v_0$ the drift velocity. Under these conditions, $\langle \dot{\mathbf{r}} \rangle = \mathbf{v}_{\boldsymbol{\kappa}}^0 + \mathbf{v}_0$ and

$$f^1 = \frac{eE_1 e^{iq_x x} e^{-i\omega t}}{\tau_{\text{ion}}^{-1} - i(\omega - v_0 q_x - v_{x,\kappa}^0 q_x)} \frac{\partial f^0(\mathcal{E}_\kappa^0)}{\partial \mathcal{E}} v_{x,\kappa}^0. \tag{S8}$$

The dynamic conductivity can now be evaluated noting that the current density is given by:

$$j_x = \frac{-e g_s g_v}{(2\pi)^2} \iint d^2\kappa \langle \dot{x} \rangle f^1, \tag{S9}$$

with $\langle \dot{x} \rangle = v_{x,k}^0 + v_0$ the mean electron velocity and $g_s g_v = 2 \times 2$ the valley and spin factors. The longitudinal conductivity (defined such that $j_x = \sigma_{q,\omega} \cdot E_1 e^{iq_x x} e^{-i\omega t}$) is evidently (for simplicity we take now the collisionless limit $\tau_{\text{ion}}^{-1} \to 0^+$):

$$\sigma_{q,\omega} = \frac{-ie^2}{\pi^2} \iint d^2\kappa \frac{\partial f^0(\mathcal{E}_\kappa^0)}{\partial \mathcal{E}} v_{x,\kappa}^0 (v_{x,\kappa}^0 + v_0) \frac{1}{\tilde{\omega} - v_{x,\kappa}^0 q_x}, \tag{S10}$$

with $\tilde{\omega} = \omega - v_0 q_x$ the Doppler shifted frequency. In the zero-temperature limit it is possible to write $\frac{\partial f^0(\mathcal{E}_\kappa^0)}{\partial \mathcal{E}} v_{x,\kappa}^0 = \frac{1}{\hbar} \frac{\partial}{\partial \kappa_x} \left[ f^0(\mathcal{E}_\kappa^0) \right] \approx \frac{1}{\hbar} \frac{\partial}{\partial \kappa_x} \left[ u(k_F - \kappa) \right]$ where $k_F$ is the Fermi wave number and $u$ is the Heaviside step function. We used the dispersion $\mathcal{E}_\kappa^0 = \hbar v_F \kappa$ of the conduction band of graphene, with $v_F$ the Fermi velocity. This approximation yields:

$$\sigma_{q,\omega} = \frac{ie^2}{\pi^2 \hbar} \iint d^2\kappa \, \delta(\kappa - k_F) \frac{\kappa_x}{\kappa} (v_{x,\kappa}^0 + v_0) \frac{1}{\tilde{\omega} - v_{x,\kappa}^0 q_x}, \tag{S11}$$

Using polar coordinates $\kappa, \varphi$ one obtains:

$$\sigma_{q,\omega} = \frac{ie^2 E_F}{\pi^2 \hbar^2} \int_0^{2\pi} d\varphi \frac{\cos^2\varphi + \frac{v_0}{v_F}\cos\varphi}{\tilde{\omega} - q_x v_F \cos\varphi}, \tag{S12}$$

where $E_F$ is the Fermi level and we took into account the linear dispersion of graphene ($\mathcal{E}_\kappa^0 = \hbar v_F \kappa$) near the Dirac cones. The integral can be evaluated analytically as

$$\sigma_{q,\omega} = \frac{+ie^2}{\hbar^2}\frac{2E_F}{\pi}\frac{1}{\tilde{\omega}}\frac{1}{A^2}\left(\frac{1}{\sqrt{1-A^2}}-1\right)\left(1+\frac{v_0}{v_F}A\right). \tag{S13}$$

with $A = \frac{q_x v_F}{\tilde{\omega}}$. Using $\frac{1}{A^2}\left(\frac{1}{\sqrt{1-A^2}}-1\right) = \frac{1}{\left(1+\sqrt{1-A^2}\right)\sqrt{1-A^2}}$ and $1+\frac{v_0}{v_F}A = \frac{\tilde{\omega}+q_x v_0}{\tilde{\omega}} = \frac{\omega}{\tilde{\omega}}$

we obtain the final and exact result for the drift-current biased graphene:

$$\sigma_{q,\omega} = \frac{+i\omega e^2}{\hbar^2}\frac{2E_F}{\pi}\frac{1}{\left(\tilde{\omega}\sqrt{\tilde{\omega}^2-(q_x v_F)^2}+\tilde{\omega}^2-(q_x v_F)^2\right)}. \tag{S14}$$

This result is fully consistent with the moving medium analogy of our earlier article [R2] as $\sigma_{q,\omega}^{\text{drift}} = \sigma_{q,\tilde{\omega}}^{\text{no-drift}}\frac{\omega}{\tilde{\omega}}$. The no-drift conductivity obtained from Eq. (S14) agrees with the result of SR and Levitov [R3, R4].

The previous analysis confirms that the Galilean Doppler shift theory is the correct answer when the e-e collisions force the electron gas to move as a whole with some drift velocity $v_0$. In such a situation, the drift effectively shifts the electrons' velocity by $v_0$, consistent with the assumption of our earlier work [R2]. In general, if the scattering is not dominated by the e-e collisions, the conductivity of the drift-biased graphene is not determined by a Galilean transformation.

We note in passing that the static current density obtained with our theory is determined from $\mathbf{j} = \frac{-eg_s g_v}{(2\pi)^2}\iint d^2\kappa\langle\dot{\mathbf{r}}\rangle f^0$ with $\langle\dot{\mathbf{r}}\rangle = \mathbf{v}_\kappa^0 - \hbar^{-1}e\partial_\kappa\left(\mathbf{E}_0\cdot\mathbf{v}_\kappa^0\tau_\kappa\right)$. After integration by parts it can be written as $\mathbf{j} = \frac{-e^2 g_s g_v}{(2\pi)^2}\iint d^2\kappa\frac{\partial f^0}{\partial\mathcal{E}}(\mathcal{E}_\kappa^0)\mathbf{v}_\kappa^0\mathbf{v}_\kappa^0\tau_\kappa\cdot\mathbf{E}_0$, which agrees with the standard theory of conduction in solids [R1].

## C. Overview of the different graphene conductivity models in the collisionless regime

Here we present a brief overview of the nonlocal conductivity of the drift-current biased graphene predicted by different models [R2-R5].

In the Svintsov and Ryzhii (SR) theory the graphene conductivity for the waves propagating parallel to the drift current is given by (restoring explicitly all the units) [R3],

$$\sigma_g^{SR}(\omega, q_x) = -ie^2 \frac{\omega}{q_x^2} \frac{1}{(\hbar v_F)^2} \Pi(\omega, q_x), \quad \text{(S15a)}$$

$$\Pi(\omega, q_x) = \frac{2}{\pi} \frac{E_F}{(1-s\beta)^2} \left( \sqrt{1-\beta^2} - \frac{s-\beta}{\sqrt{s^2-1}} \right), \quad \text{(S15b)}$$

with $E_F$ the Fermi energy, $s = \omega / |v_F q_x|$, $\beta = v_0 / v_F$, and $\omega$ in the upper-half frequency plane (UHP).

The conductivity obtained from Levitov's collisionless theory is [R4][1]:

$$\sigma_g^{Levitov}(\omega, q_x) = \frac{i\omega e^2}{\hbar^2} \frac{2E_F}{\pi} \frac{1}{\gamma(\omega - v_0 q_x)\sqrt{\omega^2 - v_F^2 q_x^2} + \omega^2 - v_F^2 q_x^2}. \quad \text{(S16)}$$

for $\omega$ in the UHP and $\gamma = 1/\sqrt{1 - v_0^2/v_F^2}$.

The real-part of the conductivity obtained from Polini's theory [R5] for the waves propagating parallel to the drift current at zero-temperature (considering only the intraband term and $\omega$ real-valued) is:

$$\mathrm{Re}\{\sigma_{g,\mathrm{intra}}^{Polini}(\omega, q_x)\} = \frac{\omega e^2}{q_x^2} \begin{cases} 2f(\omega_n, k_n)(H_{\mathrm{intra}}(A_+, B) - H_{\mathrm{intra}}(A_-, B)), & \text{if } k_n > \omega_n \\ 0, & \text{if } k_n < \omega_n \end{cases}, \quad \text{(S17)}$$

---

[1] Levitov and co-authors also introduce an alternative hydrodynamic model applicable when the electron-electron scattering rate is larger than $\omega$.

where $\omega_n = \hbar\omega/E_F$, $k_n = \hbar v_F q_x/E_F$, $f(\omega_n, k_n) = \dfrac{|E_F|}{8\pi\hbar^2 v_F^2} \dfrac{k_n^2}{\sqrt{|k_n^2 - \omega_n^2|}}$,

$A_\alpha = 2 + \alpha(\omega_n - \beta k_n)$, $B = \beta\omega_n - k_n$ and $\beta = v_0/v_F$. The function $H_{\text{intra}}(A_\alpha, B)$ is defined by:

$$H_{\text{intra}}(A_\alpha, B) = \begin{cases} G_{\text{intra}}(x_\alpha), & A_\alpha + B > 0 \text{ and } B < 0 \\ -G_{\text{intra}}(x_\alpha)\,\text{sgn}(B), & A_\alpha + B < 0 \text{ and } A_\alpha - |B| > 0, \\ 0, & \text{otherwise} \end{cases} \quad (S18)$$

where $x_\alpha = -A_\alpha/B$ and $G_{\text{intra}}(x) = x\sqrt{x^2 - 1} - \log(x + \sqrt{x^2 - 1})$.

The SR and Levitov (collisionless) theories agree exactly when the drift velocity vanishes, such that:

$$\sigma_g^{\text{no,drift}}(\omega, q_x) \approx \dfrac{i\omega e^2}{\hbar^2} \dfrac{2E_F}{\pi} \dfrac{1}{\omega\sqrt{\omega^2 - v_F^2 q_x^2} + \omega^2 - v_F^2 q_x^2}. \quad (S19)$$

Finally, the model introduced in our article [R2] predicts (considering only the intraband effects):

$$\sigma_g^{\text{Doppler}}(\omega, q_x) = (\omega/\tilde{\omega})\sigma_g^{\text{no,drift}}(\tilde{\omega}, q_x), \qquad \tilde{\omega} = \omega - q_x v_0. \quad (S20)$$

The simulations of the main text are based on the above formulas.

In our model [Eq. (S20)] the "polarizability" $\sigma_g/(i\omega)$ is transformed by the drift-current bias using the Galilean formulas $\omega \to \tilde{\omega} = \omega - q_x v_0$ and $q_x \to \tilde{q}_x = q_x$. In contrast a relativistic-type Doppler shift transformation would give instead $\omega \to \tilde{\omega} = \gamma(\omega - q_x v_0)$ and $q_x \to \tilde{q}_x = \gamma(q_x - \omega v_0/v_F^2)$ with $\gamma = 1/\sqrt{1 - v_0^2/v_F^2}$ the graphene Lorentz factor. Importantly, the no-drift polarizability $\sigma_g^{\text{no-drift}}/(i\omega)$ is transformed by the relativistic Doppler shift as:

$$\dfrac{\sigma_g^{\text{no,drift}}}{i\omega} \to \dfrac{e^2}{\hbar^2} \dfrac{2E_F}{\pi} \dfrac{1}{\gamma(\omega - q_x v_0)\sqrt{\omega^2 - v_F^2 q_x^2} + \omega^2 - v_F^2 q_x^2} = \dfrac{\sigma_g^{\text{Levitov}}}{i\omega}. \quad (S21)$$

because the terms $\omega^2 - v_\text{F}^2 q_x^2$ are "Lorentz invariant". In other words, the collisionless model by Levitov [R4] is "fully" relativistic and is essentially the result of applying a relativistic Doppler transformation to the no-drift polarizability.

From the previous discussion, it follows that Eq. (S20) and the model of Levitov will give similar results provided the Lorentz factor $\gamma$ is near unit and if in addition $\omega v_0 / v_\text{F}^2$ is negligible as compared to $q_x$. In the negative Landau damping region one has $|q_x| \sim 2\omega / v_\text{F}$, and hence provided $v_0 / v_\text{F}$ is not too large (let us say $v_0 / v_\text{F} < 0.6$) it follows that $\tilde{q}_x \approx q_x$ is typically a good approximation for $\tilde{q}_x = \gamma(q_x - \omega v_0 / v_\text{F}^2)$. Thus, Eq. (S20) typically agrees very well with the SR and Levitov models in the negative Landau damping region.

## D. Dispersion equation for the natural modes of oscillation in the graphene-dielectric-metal cavity

The dispersion equation for a system formed by a graphene sheet with drifting electrons and a plasmonic slab separated by a dielectric gap can be written as $1 - R_1 R_2 e^{-2\gamma_\text{d} d} = 0$ (here $R_1$ and $R_2$ are the magnetic-field reflection coefficients at the graphene-dielectric and dielectric-metal interfaces, respectively, $\gamma_\text{d} = \sqrt{q_x^2 - \varepsilon_\text{d} \omega^2 / c^2}$, and $d$ is the gap distance between the graphene sheet and the plasmonic slab). In the quasi-static limit $\gamma_\text{d} \approx |q_x| = q_\parallel$, and hence the characteristic equation becomes [R2]

$$1 - R_1 R_2 e^{-2 q_\parallel d} = 0 . \tag{S22}$$

The reflection coefficient at the graphene-dielectric interface is given by [R2]

$$R_1(q_x, \omega) = \frac{q_\parallel}{q_\parallel - 2\varepsilon_\text{d} \kappa_\text{g}}, \tag{S23}$$

where $\kappa_g(\omega) = i\omega\varepsilon_0 / \sigma_g^{\text{drift}}(\omega, q_x)$ and $\sigma_g^{\text{drift}}(\omega, q_x)$ is the nonlocal graphene conductivity with drifting electrons. In our model, $\sigma_g^{\text{drift}}(\omega, q_x)$ is calculated with Eq. (S20) and in the SR model with Eq. (S15).

In the quasi-static approximation, the magnetic field reflection coefficient at the dielectric-metal interface is given by:

$$R_2(\omega) = \frac{\varepsilon_m(\omega) - \varepsilon_d}{\varepsilon_m(\omega) + \varepsilon_d}. \tag{S24}$$

## *E. Incidence of an evanescent wave on a drift-current biased graphene sheet*

To illustrate the consequences of the negative Landau damping, we consider a plane wave incidence problem with the graphene sheet biased with a drift current. The incident wave is TM-polarized [see Fig. S1(a)] and is characterized by the wave number $q_x$. We focus in the case in which $|q_x| > \sqrt{\varepsilon_d}\omega/c$, which corresponds to an evanescent incident plane wave. The wave reflected by the interface is also an evanescent wave, and the superposition of the two waves generally generates a power flux towards the graphene sheet due to the material absorption. Using standard methods and the nonlocal graphene conductivity formula [Eq. (1) of the main text with the bare graphene nonlocal conductivity evaluated as in Ref. [R6], we numerically computed the *z*-component of the Poynting vector at the interface [see Fig. S1(b)].

As expected, without the drift current [blue line in Fig. S1(b)] the *z*-component of the Poynting vector is negative ($S_z < 0$) due to the material absorption by the graphene. In contrast, with the drift-current (green and purple lines in Fig. S1(b)) the *z*-component of the power flux direction may be flipped so that the energy flows *away* from the graphene sheet [R7]. This happens due to the negative Landau damping effect predicted in [R2], which enables the transfer of kinetic energy from the drifting electrons to the radiation field.

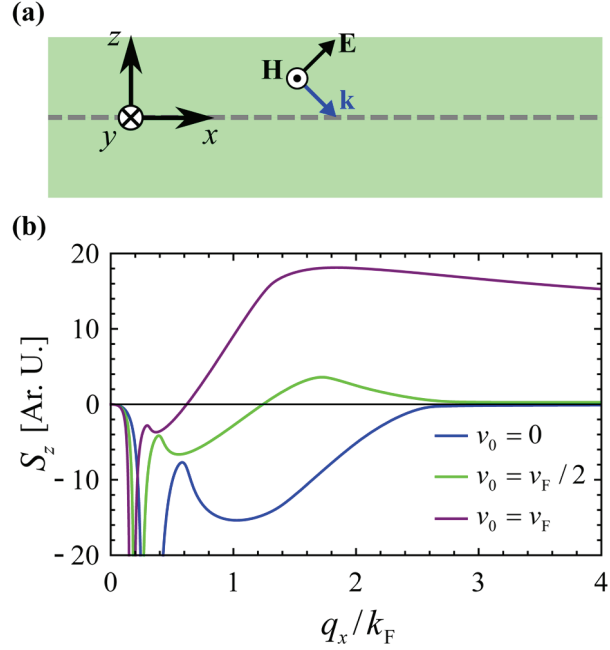

**Fig. S1.** (a) Sketch of a graphene sheet (with a drift-current bias) surrounded by a dielectric with relative permittivity $\varepsilon_d = 4$. (b) Poynting vector component perpendicular to the interface (in arbitrary units) as a function of $q_x$ for different drift velocities and $f = 15$ THz. The remaining parameters are $E_F = 0.1$ eV and $\Gamma_{intra} = 1/(0.17 \text{ ps})$.

## *F. Conductivity of a passive material in the upper-half frequency plane*

Here, we show explicitly that for a passive material (with no gain) it is necessary that $\text{Re}\{\sigma(\omega, q_x)\} > 0$ for $\omega = \omega' + i\omega''$ in the upper-half frequency plane (UHP) ($\omega'' > 0$) and $q_x$ real-valued. For simplicity, in the following we omit the dependence of $\sigma$ on $q_x$. Assuming that the conductivity is an analytic (scalar) function in the UHP it is possible to write (from Cauchy theorem):

$$\sigma(\omega) = \frac{1}{2\pi i} \int_{-\infty}^{+\infty} \frac{\sigma(\xi)}{\xi - \omega} d\xi$$

$$= \frac{1}{2\pi} \int_{-\infty}^{+\infty} \frac{\sigma(\xi)}{(\xi - \omega')^2 + \omega''^2} (\omega'' - i(\xi - \omega')) d\xi \qquad \text{for } \omega'' > 0. \qquad (S25)$$

Thus, the real-part of the conductivity in the UHP satisfies (with $\sigma = \sigma' + i\sigma''$):

$$\sigma'(\omega) = \frac{1}{2\pi} \int_{-\infty}^{+\infty} \frac{\sigma'(\xi)\omega''}{(\xi-\omega')^2 + \omega''^2} + \frac{\sigma''(\xi)}{(\xi-\omega')^2 + \omega''^2}(\xi-\omega')d\xi. \tag{S26}$$

On the other hand, the Kramers-Kronig relations imply that in the real-frequency axis [R8]:

$$\sigma''(\xi) = -\frac{1}{\pi}\text{P.V.}\int_{-\infty}^{+\infty} \frac{\sigma'(x)}{x-\xi}dx. \tag{S27}$$

This allows us to write:

$$\begin{aligned}
&\int_{-\infty}^{+\infty} \frac{\sigma''(\xi)}{(\xi-\omega')^2 + \omega''^2}(\xi-\omega')d\xi \\
&= \frac{1}{\pi}\int_{-\infty}^{+\infty} dx\,\sigma'(x)\,\text{P.V.}\int_{-\infty}^{+\infty} \frac{\xi-\omega'}{\xi-x}\frac{1}{(\xi-\omega')^2 + \omega''^2}d\xi \\
&= \frac{1}{\pi}\int_{-\infty}^{+\infty} dx\,\sigma'(x)\frac{\pi\omega''}{(x-\omega')^2 + \omega''^2}\,.
\end{aligned} \tag{S28}$$

Hence, we get simply:

$$\sigma'(\omega) = \frac{1}{\pi}\int_{-\infty}^{+\infty} \frac{\sigma'(\xi)\omega''}{(\xi-\omega')^2 + \omega''^2}d\xi, \qquad \text{for } \omega'' > 0. \tag{S29}$$

It is well-known that a passive material has $\text{Re}\{\sigma(\omega,q_x)\} > 0$ for $\omega$ and $q_x$ real-valued [R8]. Hence, the above formula implies that the same relation holds for $\omega$ in the UHP and $q_x$ real-valued.

## Supplementary information references: